\documentstyle[aps,psfig]{revtex}
\begin{document}
\draft
\title{\large \bf Empirical Formalism For 
Projectile Fragmentation and Production of \\
New Neutron-rich Nuclei with RIBs}
\author{\bf Debasis Bhowmick, Alok Chakrabarti, D. N. Basu\\
Premomoy Ghosh, Ranjana Goswami}
\address{\bf \it Variable Energy Cyclotron Centre,\\ 1/AF Bidhan Nagar, Calcutta
700 064 \\ India}
\vskip 0.9cm
\maketitle
\begin{center}
{\bf Abstract}
\end{center}
\vskip 0.9cm
	The Projectile Fragment Separator type radioactive ion beam (RIB)
facilities,
being developed in different laboratories, provide the scope for
producing many new exotic nuclei through fragmentation of
high energy radioactive ion (RI)
beams. A new empirical parametrization for the estimation of
cross-sections of projectile fragments has been prescribed for
studying the advantages and limitations of high energy RI beams for the
production of new exotic nuclei. The parametrization reproduces the
experimental data for the production of fragments 
from neutron rich projectiles accurately in contrast 
to the existing parametrization which tends to overestimate the 
cross-section of neutron rich fragments in most cases. 
The modified formalism has been used to
compute the cross-sections of neutron-rich species produced by
fragmentation of radioactive projectiles (RIBs). It has been found that,
given any limit of production cross-section, the exoticity of the
fragment increases rather slowly and shows a saturation tendency as the
projectile is made more and more exotic. This essentially limits, to an
extent, the utility of very neutron-rich radioactive beams vis-a-vis
production of new neutron-rich exotic species.
\vskip 5cm
PACS NO : 25.60.Dz, 21.10.Dr 

\newpage

\section{\bf Introduction}

\vskip 0.5cm
	Synthesis of many new $\beta$-unstable nuclei
and production of such nuclei with appreciable
yield for structure studies and other experiments are becoming feasible
with the advent of Radioactive Ion Beam (RIB) facilities and related
activities in many laboratories [1], all over the world, for
different kinds of facilities following different routes of nuclear
reactions.

	Existing high energy heavy ion accelerators provide the
opportunity for producing $\beta$-unstable nuclei
through Projectile Fragmentation (PF) reactions
with stable projectiles. Projectile fragments Separator (PFS) technique,
as adopted by different laboratories [2] provides RIBs of high energies
suitable for production
of more exotic nuclei through PF reaction of secondary unstable
projectiles. Reliable estimation of yields of
these  exotic nuclei produced by fragmentation of secondary RI
projectiles is very important for assessing the advantages
and limitations of RIBs towards production of new 
exotic species.

	Estimation of the production cross-sections new exotic
species, especially of the n-rich nuclei, through PF reactions of 
RIBs is our topic of interest for this
article. It is worth mentioning that similar exercises on production of
exotic nuclei with RIBs through multinucleon transfer reactions
and CN-evaporation reactions, the other two important
reaction mechanisms for production of exotic nuclei, have already been
addressed [3].  

	In this article we have attempted to develop an empirical
formalism for the estimation of fragments' cross-sections from RI
projectiles and for this purpose we started with the well studied target
fragmentation processes. Target fragmentation cross-sections have been
measured for various experimental conditions. 
Empirical parametrization to fit the experimental mass and charge yield
$\sigma$(A,Z) was initiated by Rudstam [4] and has thereafter been
elaborated by many others [5,6]. The formalism of Summerer et al [6] is
now widely used to predict the cross-sections of projectile fragments in
PF reactions[7], since this reaction can be considered, in the projectile
rest frame, in the same footing as that of target fragmentation in the
target rest frame.

	The parametric expression of Summerer et al can reproduce the
experimental cross-sections of projectile fragments quite accurately    
only if the fragments are not too   
exotic and the projectile is the most $\beta$-stable one, e.g. $^{40}$Ca
instead of say $^{48}$Ca. This can be clearly seen from Fig. 1, where 
the cross section data [8] for ${}^{21-24}$O, ${}^{23-25}$F, ${}^{25,29}$Ne,
${}^{29-35}$Na, ${}^{34-38}$Mg, ${}^{35-41}$Al, ${}^{41-45}Si$ from $^{48}Ca
+ ^{181}Ta$ at 70 A MeV reaction are represented along with the
estimations by the formalism of Summerer et al. It can easily be seen
that the formalism overestimates the cross-sections of exotic (n-rich)
fragments. Hence this formalism can not be used to predict reliably the
cross-sections of even more exotic n-rich fragments produced in 
PF reactions of n-rich exotic projectiles (RIBs).   
One thus needs to develop a modified empirical formalism if one aims at
a fairly accurate prediction of the cross-sections of new n-rich nuclei
which can hopefully be produced with n-rich radioactive projectiles.

The organization of the paper is as follows : Section II of this paper 
has been devoted for a brief description of the empirical
parametrizations, and the modifications that have been been carried out
in this work; results and conclusions are discussed in sections III and IV
respectively.
\vskip .5cm
\section{\bf Empirical Parametrization of Fragmentation Cross-sections.}
\vskip .3cm

	The analytical formulae for the yield distribution as a 
function of the fragment mass and charge
$\sigma (A,Z)$ is conventionally written as, 
\begin{equation}
\sigma (A,Z) = [Y(A)].[n exp(-R|Z_P - Z|^U)].
\end{equation}
The first term Y(A) represents the mass yield, i.e. the sum of
the isobaric cross-sections with mass A, while the second term describes the
charge dispersion around its maximum $Z_{\it P}$. R is the width parameter,
which actually controls the shape of the charge dispersion and U
its exponent. The factor  n simply serves to normalize the integral to
unity.

	Abu-Magd, Friedman and Hufner [9] have shown that for
relativistic proton induced reactions one obtains the functional form of
Y(A) from multiple scattering and an approximation to the evaporation
chain for spallation as :
\begin{equation}
Y(A) = \sigma _{R}P(A_{P}).exp[-P(A_{P})(A_{P} - A_{F})],
\end{equation}
where the best fitted slope paremeter $P(A_{P})$ is written as :
\begin{equation}
ln P(A_{P}) = -7.57\times 10^{-3}A_{P} - 2.584.
\end{equation}
The factor $\sigma_{R}$ is chosen as the sum of the target and
projectile radii rather than the square of the sum as : 
\begin{equation}
\sigma_{R} = 450(A_{P}^{1/3} + A_{T}^{1/3} - 2.38 ),
\end{equation}
where $A_{P}$ and $A_{T}$ denote the projectile and target mass numbers
respectively.

	Furthermore, in order to get the distribution of nuclear charge,
$Z$, for a given fragment mass number $A$., three parameters (in Eqn.1) R,
$Z_{\it P}$ and U must be known. Since these three parameters are strongly
correlated, it is really difficult to have unique estimation with a
least-square fitting technique and hence people have chosen instead to fix
first the exponent U and then best fitted values of $Z_{\it P}$ and R were
obtained by fitting the data.

	It is important to mention that all the presently available data
indicates that Rudstam's early suggestion, U = 1.5 gives a very good
description in the n-rich side of the distribution,
whereas the p-rich side of the distribution falls off like
a Gaussian with U=2. But for strictly symmetric charge dispersion, the
values of U may be taken in between 2 and 1.48.

The width parameter R  which is a function of fragment 
mass only, irrespective of the projectile nucleus, has been approximated 
with an exponential of the form as :
\begin{equation}
lnR(A_{F}) = -6.770\times 10^{-3}A_{F} + 0.778.
\end{equation}
\vskip 0.3cm

	In general, the parametrization of $Z_{\it P}$ is very important.
Several parametrization have already been performed and from their
analyses for the fragmentation cross-sections the conclusions can be
summerized as follows :
\begin{itemize}
\begin{enumerate}
\item For stable projectile, the maximum of the charge distributions are always on the
neutron-deficient side of the valley of $\beta$-stability.

\item For projectile close to $\beta$-stability $Z_{\it P}$ is a function of
the fragment mass only.

\item For more n-rich or p-rich projectile the fragments remember
the  neutron or proton excess of the projectile to varying extents (the 
so called memory effect). 
\end{enumerate}
\end{itemize}
Following the idea of Chu et al [10], (where $\Delta$ has been considered
to take into account the n- or p-richness of the fragments) Summerer et
al  has finally landed up with the following expression of $Z_{\it P}$ as
:
\begin{equation}
Z_{\it P}(A_{F}) = Z_{\beta}(A_{F}) + \Delta + \Delta_{m},
\end{equation}
where $Z_{\beta}(A_{F})$ is approximated by a smooth function [11] as :
\begin{equation} 
Z_{\beta}(A_{F}) = A_{F}/(1.98 + 0.0155 A_{F}^{2/3})
\end{equation}
and the differences $\Delta$ and $\Delta_{m}$ have been parametrized as,
\begin{equation}
\Delta = 2.041\times A_{F}^{2}\times 10^{-4} \ \ \ \ \ \ \  A_{F}\le 66,
\end{equation}
\begin{equation}
\Delta = 2.703\times A_{F}\times 10^{-2}-0.895 \ \ \ \ \  A_{F}\ge 66,
\end{equation}
\begin{equation}
\Delta_{m} = 
[C_{1}(A_{F}/A_{P})^{2} + C_{2}(A_{F}/A_{P})^{4}]\Delta_{\beta}(A_{P}),
\end{equation}
with
\begin{equation}
\Delta_{\beta}(A_{P}) = Z_{\it P} - Z_{\beta}(A_{P}).
\end{equation}
$C_{1} = 0.4$, $C_{2} = 0.6$ are assigned for n-rich projectile, 
while for p-rich
$C_{1} = 0.0$, $C_{2} = 0.6$. 
The factor $\Delta_{m}$ was introduced to take care of the 'memory
effect', that is the effect of the exoticity of the projectile on the
fragment production cross-sections. However, it is more appropriate and
aesthetically appealing to introduce exoticities of the projectile and
the fragment directly in the expression for the charge distribution
instead of the parameter $\Delta$ and $\Delta_{m}$ to take care of the
memory effect.

Keeping in mind that the drip line is closer to the $\beta$-stability line
at lower $Z$ and moves more and more away as $Z$ increases, we define
the exoticity of the projectile and fragment as, 
\begin{equation}
\rho_{F}
=\frac {A_{F} - A_{F}^{s}}{Z^{F}} 
\end{equation}
and 
\begin{equation}
\rho_{P} = \frac {A_{P} - A_{P}^{s}}{Z^{\it P}},
\end{equation}
        where $Z^{F}$ and $Z^{P}$ are the atomic numbers of the
fragment and the projectile respectively. $A_{F}(A_{P})$ is the mass
number of the fragment (projectile) and $A_{F}^{s}(A_{P}^{s})$ is the
mass number of the $\beta$-stable isotope corresponding to
$Z^{F}(Z^{P})$.

	$\rho_{F}$ and $\rho_{P}$ are added in the expression of most
probable charge $Z_{P}$ as 0.8$\rho_{F}-2.\rho_{P}$, that is  with opposite
signs. This is because more exotic projectile should favour production of more
exotic fragments whereas the production cross-section of fragments from
a given projectile should decrease as the fragment becomes more exotic.

	We have taken care of the 'memory effect' through the exoticity
parameters $\rho_{P}$ and $\rho_{F}$. However, to have the centroid
perfectly at $A_{F} = A_{F}^{s}$ for $A_{P} = A_{P}^{s}$ we have added
the factors $\Delta^{s}$ and $\Delta_{m}^{s}$ in the expression for the
most probable charge $Z_{P}$, where $\Delta^{s}$ and $\Delta_{m}^{s}$
are the values of $\Delta$ and $\Delta_{m}$ on the $\beta$-stable line
and are calculated from equations (8), (9) and (10) by substituting $A_{F}
=A_{F}^{s}$ and $A_{P} = A_{P}^{s}$. For $Z\le 40$, $A_{F}^{s}$ 
and $A_{P}^{s}$ are calculated using the expression : 
\begin{equation}
A^{s} = 2.08Z + 0.0029{Z}^{2} + 7.00\times 10^{-5}{Z}^{3},
\end{equation} 
which is a slightly modified form of the expression used by Charity et
al [12] earlier.

	To have a right match with the experimental data 
the width parameter R and the exponent U have been adjusted in our
formalism as,  
\begin{equation}
ln R = -6.770\times 10^{-3}A_{F} + 24.11\times
10^{-2}{(A_{P}-A_{F})}^{1/3}
\end{equation}
and U = 1.57 instead of 1.5 used by Summerer et al for n-rich 
fragment while for p-rich side U=2, with the normalisation factor
$n=\sqrt{\frac{R}{\pi}}$. 

	The $(A_{P}-A_{F})^{1/3}$ dependence of the
width parameter R is a new addition and 
has been introduced since intutively the isospin
fluctuation should become less and less probable as bigger and bigger
chunks/portions of the projectile are removed from the projectile. These
implies that the width $(R^{-1})$ of the charge distribution should
decrease as the fragment becomes more and more lighter as compared to
the projectile.

	To take care of the fact that projectiles beyond the n-drip line
are unrealistic and can not therefore lead to an enhancement of
cross-section for the production of exotic species, we have introduced 
a factor, 
\begin{equation}
\xi^{d} = 1.22\times
\frac {A_{P}^{d} - A_{P}^{s}}{Z^{\it P}}
\end{equation}
in the expresion of the most probable charge $Z_{P}$. $A_{P}^{d}$ is the
mass number at the drip line corresponding to the projectile's charge
number $Z^{P}$. $A_{P}^{d}$'s are decided on the basis of the mass
formula by Janeeke-Masson [13]. The factor $\xi^{d}$ has been considered
only for cases where the projectile is beyond the n-drip line. It is
important to note $\xi^{d}$ is, but for the constant factor 1.22, just
the value for the exoticity at the drip line corresponding to $Z^{P}$.

        The same expression and parametrization [equations (2), (3) and
(4), ref. 6] for the mass yield Y(A) has been used in our formalism.
Thus the target dependence of the fragmentation cross-section has been
taken care only through equation (4).	

	The final form of our expression thus becomes,
\begin{equation}
\sigma (A,Z) =
Y(A).n.exp[-R|Z_{\beta}+\Delta^{s}+\Delta_{m}^{s} + 0.8\rho_{F}-2.\rho_{P}-\xi^{d}-Z|^{U}]
\end{equation}
with Y(A) given by eqn.(2).
\vskip 0.5cm
\section{\bf Results.}
\vskip 0.1cm
        Fig.1 shows a comparison of the modified parametrization with the
older one along with a number of experimental values of cross-sections
of various fragments with different exoticities for the nuclear reaction
: $^{181}Ta(^{48}Ca,X), E(^{48}Ca) =$ 70 A MeV (ref.8). The dotted
lines show the results of the earlier [6] parametrization while the 
solid zig-zag lines join the experimental data points. The modified 
parametrization has been shown by solid lines. 
The modified relationship gives excellent match with that of the
experimental data and obviously a much better match compared to the
earlier 
one which is depicted in Fig.1. A better fit has always been obtained
for more exotic species. However, as has already been
discussed, the earlier parametrization as well as the present one give
similar results for less exotic fragments produced from 
less exotic projectile. This feature
is clearly visible in Fig.2, where estimated 
cross sections for various fragments 
produced in the nuclear reaction : $^{181}Ta(^{50}Ti,X),
E(^{50}Ti) =$ 80 A MeV. [14] are shown along with the experimental data.

	In Fig.3. the results of reactions $^{9}Be(^{50}Ti,x)$, $E(^{50}Ti)
= 80 A MeV$ are fitted, that is, for the same projectile as in Fig.2 but
with $^{9}Be$ target instead of $^{181}Ta$.
It can be seen from figures 2 and 3 that our formalism can fit the
existing data very well for both $^{181}Ta$ and $^{9}Be$ target.
	In Fig.4. we have shown the full specturm of the predicted
production cross-sections
of Si fragments for three different Ca
projectiles viz. $^{42}Ca, {}^{50}Ca, {}^{58}Ca$.  A comparison of the
earlier parametrization (dotted lines) with the present one (solid
lines)is shown here. It can be seen  that the earlier parametrization
tends to  
overestimate in the n-rich side and the extent of the overestimation
increases as the projectile or the fragment of interest become more
exotic.

	To illustrate the point further and in a more quantitative
fashion, we compare the ratio $\frac{\sigma_{old}}{\sigma_{new}}$, where
$\sigma_{old}$ and $\sigma_{new}$ are the cross-sections for fragment
production as calculated on the basis of the earlier and the present
parametrization respectively, for two cases :
\begin{itemize}
\begin{enumerate}
\item{production of $^{35}$Si and $^{40}$Si from the projectile
$^{50}$Ca}
\item{production of $^{40}$Si from projectiles $^{50}$Ca and $^{58}$Ca}.
\end{enumerate}
\end{itemize}
In the first case the extent of overestimation (the ratio) increases
from a factor of 3 for $^{35}$Si to a factor of 35.5 for the more exotic
fragment $^{40}$Si. For the second case the factor increases from 35.5 in
the $^{50}$Ca projectile case to about 52.2 in the case of more exotic
projectile $^{58}$Ca. Thus the discrepancy between the earlier and the
present formalism becomes very serious in the RIB case where one needs
to address the question of production of very exotic species from exotic
projectiles.

	In Fig.5 the maximum exoticity of the fragments (Si) are plotted
against the corresponding projectile's (Ca) exoticity for two different
cross-section limits viz. 0.01 mb and 1.0 mb. That is to say, given the
lower limit of cross-section the exoticity of the maximum exotic Si
isotope that can be produced from the fragmentation of a given Ca
isotope is plotted as a function of exoticity of the Ca isotopes.
Results for both the earlier and the present formalism are shown. It is
evident that with the relaxation of the cross-section limit more exotic
fragments are produced which is expected, although, the n-drip line 
is ($\rho_{F}$ = 0.71 for Si)
never reached with 0.01 mb cross-section limit. Calculation with present
parametrization shows that the n-drip line can be reached with
cross-section in the range $0.001\le\sigma\le 0.01$ mb for projectile
exoticities greater than 0.65, which means drip-line for Si
can be reached only with $^{56}$Ca or more exotic Ca projectile.
Interestingly, as the projectile is made more and more exotic, the
exoticity of the fragment increases and ultimately shows a saturation
tendency. The flat portion for projectile exoticities beyond the drip
line (for Ca, $\rho_{P}$=0.75 at drip line which corresponds to
$^{58}$Ca ) is a consequence of the factor $\xi^{d}$ in the charge
distribution. This tends to limit 
the usefulness of going for very neutron rich projectile beams.
The earlier formalism predicts a much steeper increase of the fragment
exoticity with the projectile exoticity and thus overestimates
as compared to the present formalism, the 
advantage of RI projectiles for the production of new n-rich species.
 
	The main interest of the present work has been to develop a
formalism which can be used to predict reliably the production
cross-sections of n-rich products from n-rich RIBs. However, to see how
well the present formalism can predict cross-sections
for the production of p-rich nuclei
we have also considered the systems $^{58}Ni(^{78}Kr,X)$,
$E(^{78}Kr)$ = 75A MeV [15] and $^{27}Al(^{86}Kr,X)$, $E(^{86}Kr)$ = 70A MeV
[16], for which the experimental cross-section values for a large
number of p-rich nuclei are available in the
range $Z=30$ to $Z=38$ and $Z=33$ to $Z=39$ respectively. 
The experimental data (solid points) of p-rich nuclei, the
results of Summerer et al (dotted lines) and the results of the present
formalism (solid lines) are shown in figures 6 and 7.

	In the $^{78}Kr$ projectile case (Fig.6), calculations based on
Summerer et al formalism give a slight overall better match with the
experimental cross-sections as compared to predictions based on the
present formalism. However, in the case of $^{86}Kr$ projectile (Fig.7),
the present formalism reproduces the experimental data much better than
the predictions of Summerer et al. It is important to note that
$^{78}Kr$ is rather a p-rich projectile in contrast to $^{86}Kr$, which
is slighly n-rich (although both are $\beta$-stable isotopes of Kr).
This, together with the results in case of rather n-rich
projectiles $^{48}Ca$ and $^{50}Ti$, show that, while our 
formalism can predict 
accurately the cross-sections of fragments produced from n-rich
projectiles, the predictions are not so accurate for fragments produced
from p-rich projectiles.

	In case of $^{86}Kr$ projectile our formalism can not predict
accurately the cross-sections of Rb and Sr isotopes, although the
matching is much better as compared to the predictions of the earlier
formalism. Especially in the
case of Sr-isotopes the difference between the prediction and the
experimental data is more than an order of magnitude. However, we note
that production of Sr-isotopes
from Kr-projectile involves two proton pick-up
reactions and therefore, the production cross-sections should have a 
significant component from
the pick-up reaction. This might explain the poor matching in the case of
Sr-isotopes.	

	To examine how well our formalism can reproduce the experimental
data at higher energies we have plotted in Fig. 8, the experimental
cross-sections of isotopes of different elements from Ca to Kr, produced
in 500 A Mev $^{86}$Kr [17] along with the estimations based on the present
formalism as well as the earlier formalism.

	In order to have a fairly good agreement with the experimental
data we need to use a value of the exponent U of the charge distribution
equal to  1.4 instead of 1.57 which was used in the lower energy ($\le$
100 A MeV) domain. As is evident from Fig.8, the agreement with the
experimental data is not as excellent as it has been the case at
energies below 100 A MeV.

	At present no experimental data is available for the production
cross-sections of very n-rich exotic fragments from $^{86}$Kr
projectile. The availability of such data would have allowed us to check
the accuracy of the predictions of the present formalism for projectiles
upto $A \sim$100. The present formalism predicts a value of 7pb, 2.2pb
and 180 pb for the cross-section of $^{78}$Ni produced by the
fragmentation of $^{86}$Kr, $^{98}$Kr and $^{86}$Ge respectively. The
cross-section values clearly indicate that the use of a more exotic
projectile leads to higher cross-section for production only when its mass
number is not far away from the fragment's mass number. This is an
artefact of the ${(A_P - A_F)}^{1/3}$ dependence of the width ($R^{-1}$) 
which tends to compensate the
gain in the cross-section due to the increase in exoticity of the projectile by the
corresponding loss due to decrease in the width. 

\section{\bf Conclusion.}
\vskip 0.3cm 

	In this article we have attempted to develop an empirical
formalism for estimating the cross-sections of
exotic fragments produced from exotic projectiles (RIBs). In the present
formalism we have assumed following earlier prescriptions the
factorizability of the production cross-section. The charge distribution
part of the earlier formalism [6] has been throughly modified in the
present work, while the expression for the mass yield has been kept
unchanged. 

	It has been
shown that the present (modified) formalism can reproduce accurately
the recent experimental data for the production of a number of
neutron-rich species from $^{48}$Ca and $^{50}$Ti projectiles. The
present formalism can, therefore, be used to predict much more realiably
the production cross-section of new n-rich nuclei produced by
fragmentation of n-rich radioactive projectiles.

A number of physically appealing and transparent new parameters are
introduced in the expression for charge distribution in our formalism.
The introduction of exoticity parameters for the projectile and the
fragment to take care of the so called 'memory' effect is one of them.
The excellent matching with the experimental data (Fig.1 and 2) for
n-rich exotic fragments justifies further their inclusion. Moreover, the
representation in terms of the exoticity of the fragment and the
projectile clearly brings out how much one can expect to gain interms of
production of more exotic species by using a more exotic projectile. The
fact that the slope of this curve is rather flat irrespective of the
lower limit of the cross-section one has considered, tends to offset
some of the advantages of using very n-rich projectiles for the
production of new n-rich nuclei.

	The introduction of $(A_{P}-A_{F})^{1/3}$ dependence of the
width parameter R represents another new physics input which has been
introduced to take care of the intuitive expectation that the width of
the charge distribution should depend on the fraction of the projectile
that has been chopped out in the fragmentation reaction. 
This dependence is crucial for having an uniform good matching with the
experimental cross-sections over a wide range of fragments produced from
the same projectile, e.g. in the case of production of different isotopes
of all elements between oxygen and silicon from the fragmentation of
$^{48}$Ca (Fig.1). Furthermore, this factor tends to offset the
advantage (higher production cross-section) of using more and more
exotic projectiles of a given atomic number for the production of a
given n-rich exotic fragment of interest.

	Although, the present formalism results in a much better overall
fit with the experimental data for a varities of projectiles and
projectile energies as compared to the earlier formalism [6], the
agreement is not equally good in all cases. For example, the agreement
is not very satisfactory for fragments produced from $^{78}$Kr.
The same is true for fragments produced from the fragmentation of
$^{86}$Kr at 500 A MeV. 
While a better agreement in these cases would
certainly have been a more desirable feature, 
the main interest of the present work has been to develop an
empirical formalism which can be used to predict reliably the production
cross-sections of n-rich nuclei produced from fragmentations of n-rich
radioactive projectiles. The present formalism is expected to
be quite good for the said purpose. 
 
\newpage
\begin{figure}
\psfig{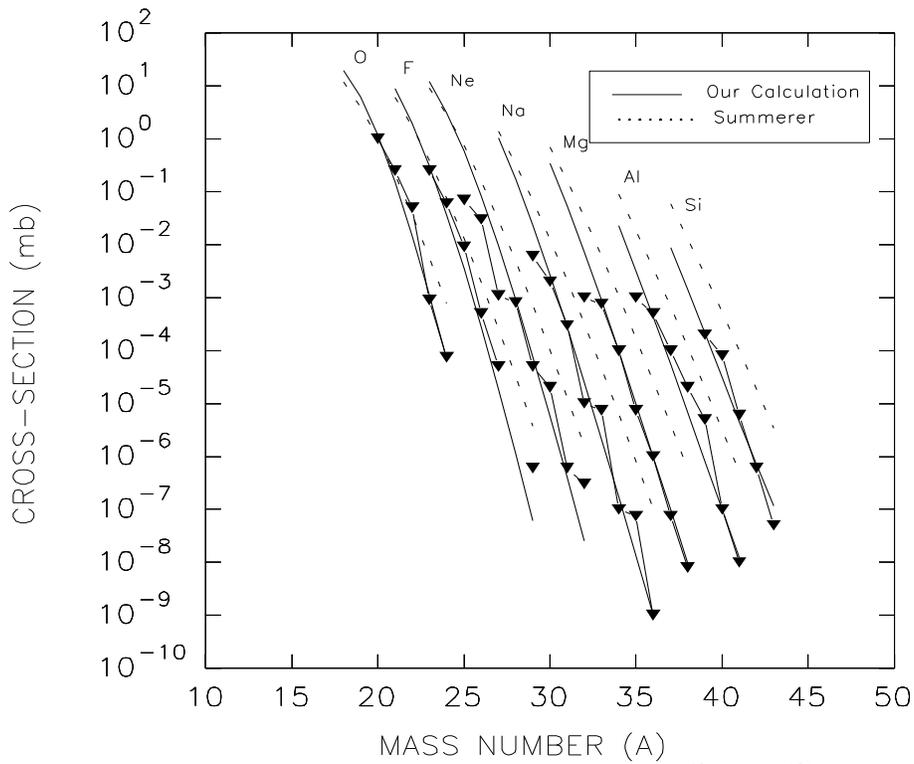}
\caption{Production cross-sections of projectile fragments of $^{48}$Ca
for ${}^{181}$Ta target, compared with those calculated by the old and
modified parametrizations.}
\end{figure}
\begin{figure}
\psfig{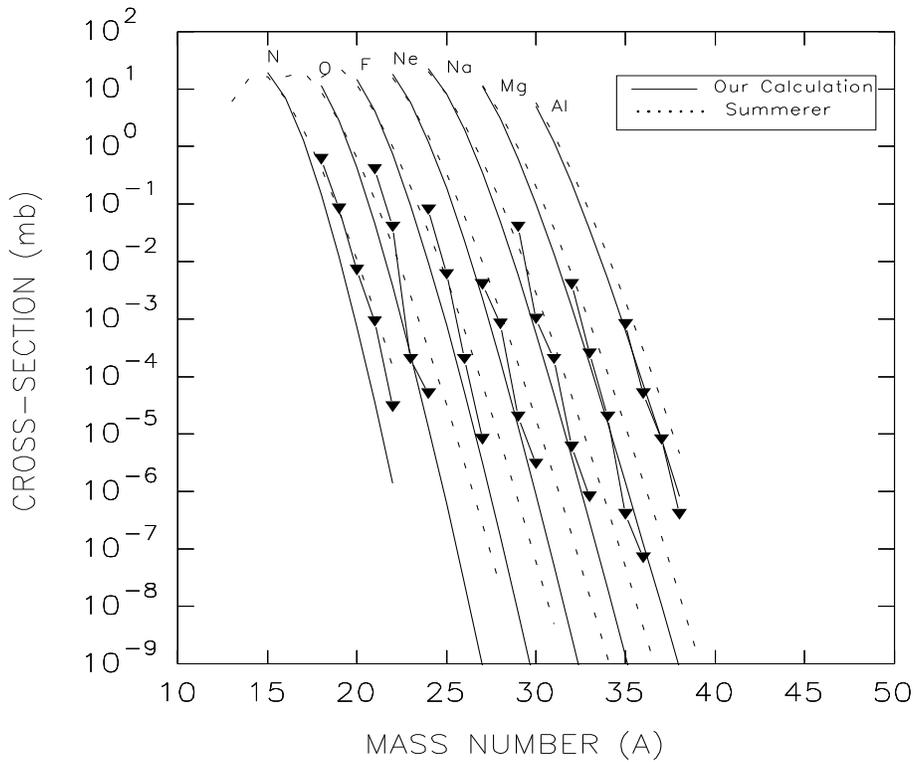}
\caption{ same as Fig.1.with $^{50}Ti$ as projectile.}
\end{figure}
\newpage
\begin{figure}
\psfig{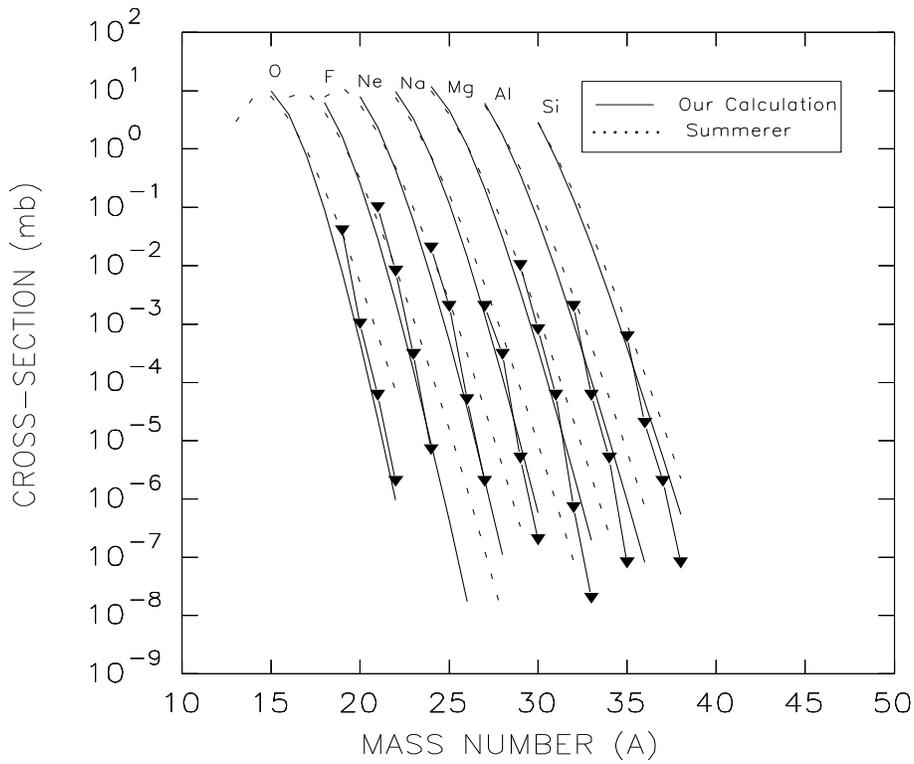}
\caption{ same as Fig.2.with the same projectile but $^{9}Be$ as target}
\end{figure}
\begin{figure}
\psfig{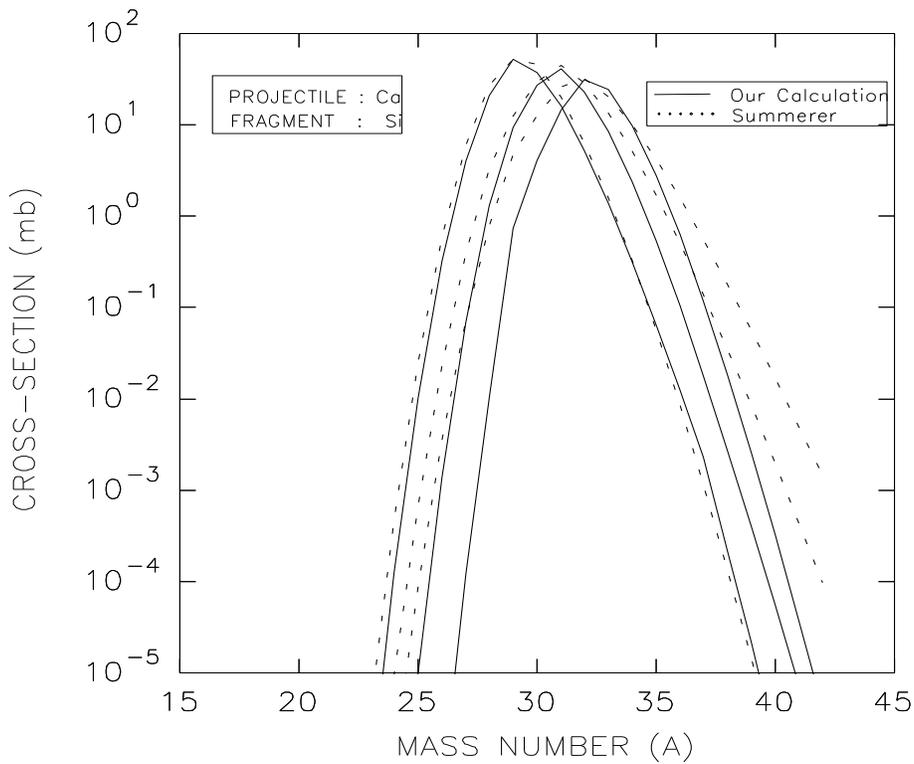}
\caption{Production cross-sections of Si isotopes, calculated from the
old and the modified parametrization for three different Ca
projectiles viz.$^{42}Ca, ^{50}Ca, ^{58}Ca$.}
\end{figure}
\newpage
\begin{figure}
\psfig{figure=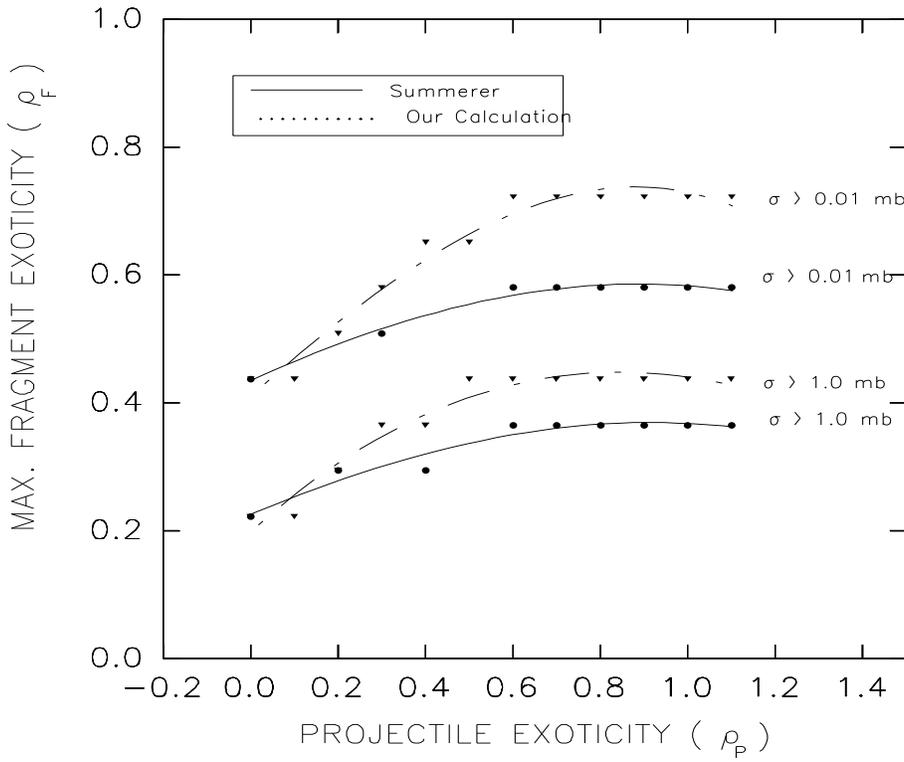,height=10cm,width=12cm}
\caption{The variation of the maximum fragment exoticity($\rho_{F}$)
with the exoticity  of the projectile for the old and the modified
formalism in the n-rich side.} 
\end{figure}
\newpage
\begin{figure}
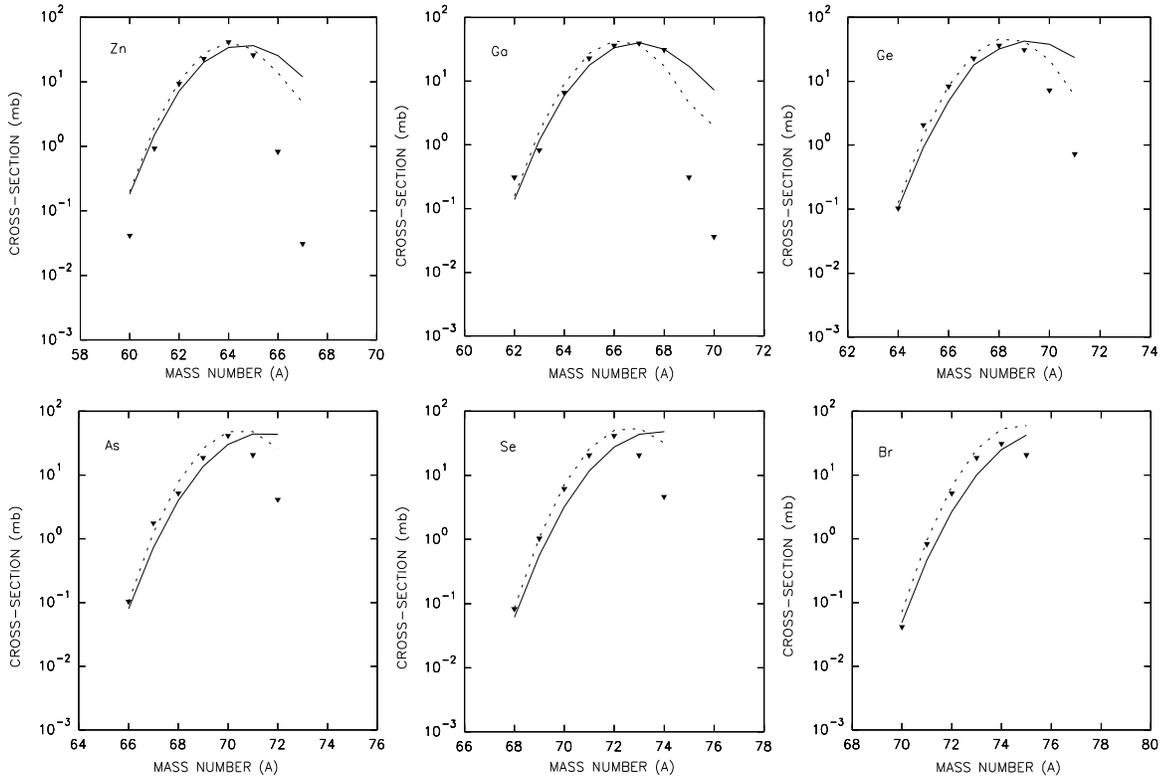

\begin{tabular}{ccc}         
   \psfig{file=msu1.ps,width=5cm,height=5cm}  &
   \psfig{file=msu2.ps,width=5cm,height=5cm}  &
   \psfig{file=msu3.ps,width=5cm,height=5cm}  \\
   \psfig{file=msu4.ps,width=5cm,height=5cm}  &
   \psfig{file=msu5.ps,width=5cm,height=5cm}  &
   \psfig{file=msu6.ps,width=5cm,height=5cm}  \\
\end{tabular}
\caption{Isotopic cross-section for the elements between Zinc and Bromine from
the reaction $^{78}Kr + ^{58}Ni$ at 75 A MeV. The solid points indicate
the measured production cross-sections. The dotted lines represent the
parametrization by Summerer et al, while the solid line represent our
calculation.}
\end{figure}
\newpage
\begin{figure}
\begin{tabular}{ccc}         
   \psfig{file=msul1.ps,width=5cm,height=5cm}  &
   \psfig{file=msul2.ps,width=5cm,height=5cm}  &
   \psfig{file=msul3.ps,width=5cm,height=5cm}  \\
   \psfig{file=msul4.ps,width=5cm,height=5cm}  &
   \psfig{file=msul5.ps,width=5cm,height=5cm}  &
   \psfig{file=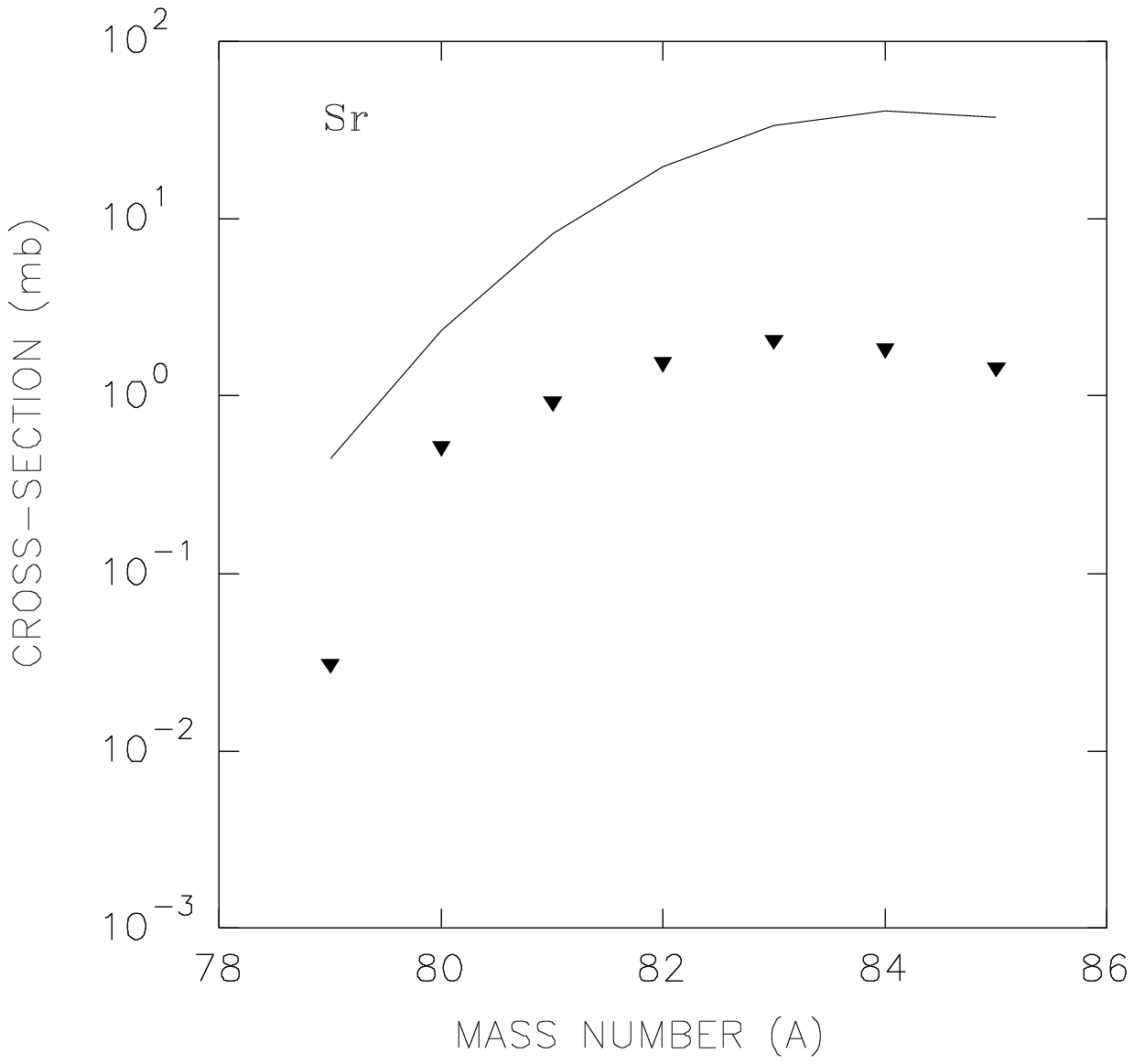,width=5cm,height=5cm}  \\
\end{tabular}
\caption{Similar to Fig.6, but for different reaction : $^{86}Kr 
+ ^{27}Al$ at 70 A MeV.} 
\end{figure}
\begin{figure}
\begin{tabular}{ccc}         
   \psfig{file=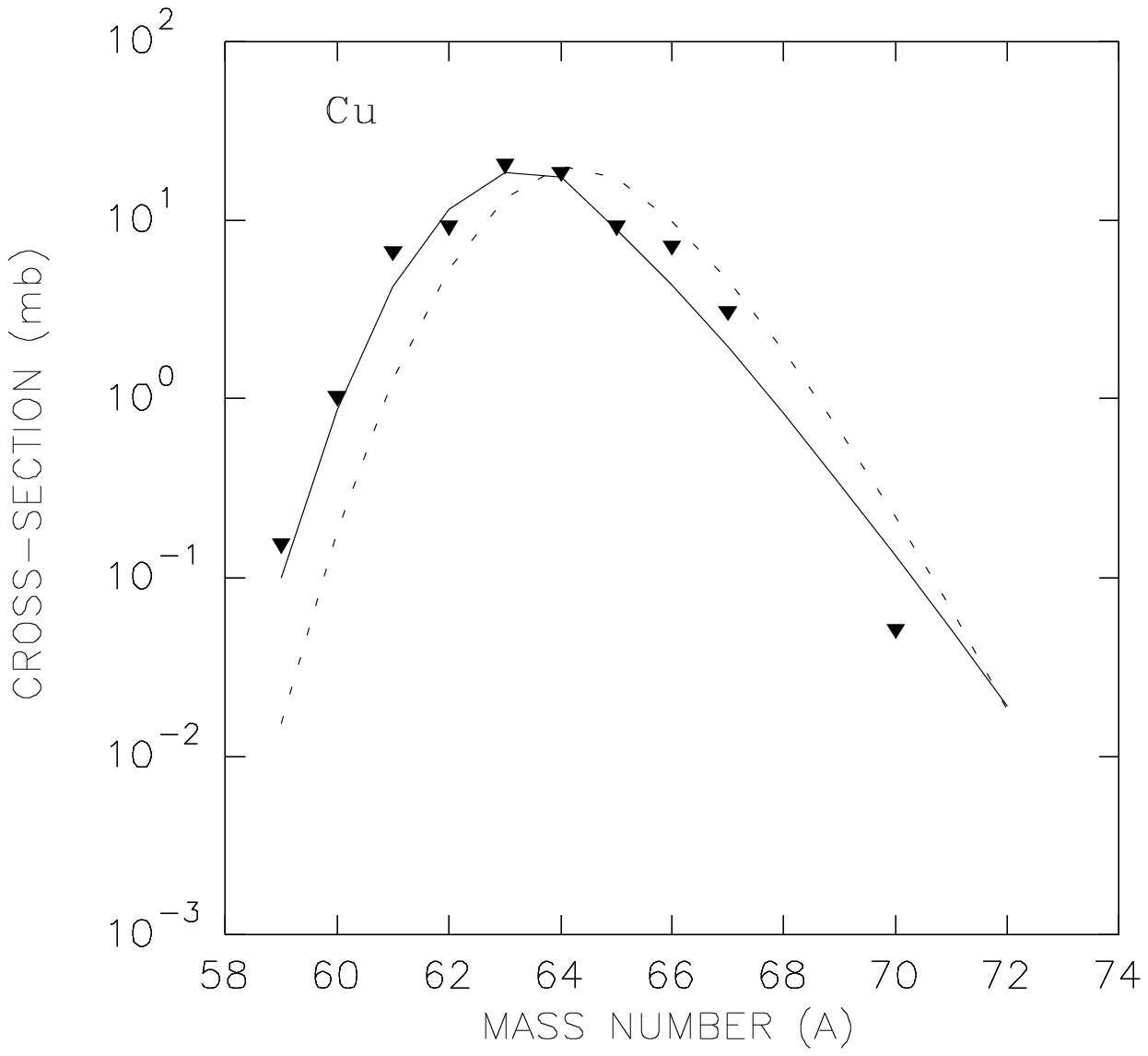,width=5cm,height=5cm}  &
   \psfig{file=gsi22.ps,width=5cm,height=5cm}  \\
   \psfig{file=gsi33.ps,width=5cm,height=5cm}  &
   \psfig{file=gsi44.ps,width=5cm,height=5cm}  \\
   \psfig{file=gsi55.ps,width=5cm,height=5cm}  &
   \psfig{file=gsi66.ps,width=5cm,height=5cm}  \\
   \psfig{file=gsi77.ps,width=5cm,height=5cm}  &
   \psfig{file=gsi8.ps,width=5cm,height=5cm}   \\
\end{tabular}
\caption{Isotopic cross-sections for elements between copper (Z=29) and
Krypton (Z=36) from the reaction : $^{86}Kr + ^{9}Be$ at 500 A MeV. 
The solid points indicate the measured production cross-sections. 
The dotted lines represent the parametrization by Summerer et al, 
while the solid line represent our calculations.}
\end{figure}
\newpage
\bf{References:}

\begin{itemize}
\begin{enumerate}
\item J.M. D'Auria, NIM, B99, 330 (1995)
\item  A. C. Mueller et al, NIM, B56/57, 559 (1991) and referrences
therein;  B. M. Sherrill et al, NIM, B56/57, 1106 (1991);
H. Geissel et al, GSI Scintific Report 1988, Report No. GSI-89-1,
277, 1989 and references therein. 
\item C. H. Dasso et al, Phys. Rev. Lett 73, 1907 (1994); P.Ghosh et al,
comm. in Phys. Lett. B. 
\item  G. Rudstam, Z. Naturforsch. 21a, 1077 (1966)
\item  R. Silberberg, C. H. tsao, Ast. Jour. Supp. 220(II),25,335 (1973)
\item  K. Summerer et al, Phys.Rev. C42, 2546 (1990)
\item  J. A. Winger et al, NIM B70,380 (1992); D. Guillemaud-Mueller et
al, Phys. Rev. C41,937 (1990). 
\item  M.Notani et al, RIKEN Accel, Prog, Rep, 30, 48 (1997)
\item  A. Y. Abu-Magd et al, Phys. Rev. C34, 113 (1986)
\item Y.Y. Chu et al, Phys. Rev. C4, 2202 (1971); $\it ibid$. Phys. Rev.
C10, 156 (1974).
\item P. Marmier and E. Sheldon, Physics of Nuclei and Particles
(Academic, New York and London, 1971) volI. p,15.
\item R. J. Charity et al, Nucl.Phys. A476, 516 (1988)
\item Atomic Data and Nuclear Data Tables, Vol.39, No.2, p296 (1988)
\item M. Nontani et al, RIKEN. Accel. Prog. Rep. 29, 50 (1996)
\item R. Pfaff et al, Phys. Rev. C53, 1753 (1996)
\item R. Pfaff et al, Phys. Rev. C51, 1348 (1995)
\item M. Weber et al, Nucl. Phys. A578, 659 (1994)
\end{enumerate}
\end{itemize}

{\bf Acknowledgements :}

One of the authors (DB) wishes to acknowledge the financial support from
CSIR, Govt. of India.
\end{document}